# Spin-polaron band of heavy carriers in the heavy-fermion ferromagnetic superconductor UGe$_2$

Vyacheslav G. Storchak[1], Jess H. Brewer[2], Dmitry G. Eshchenko[3], Patrick W. Mengyan[4], Oleg E. Parfenov[1] & Dmitry Sokolov[5]

[1]*Russian Research Centre "Kurchatov Institute", Kurchatov Sq. 1, Moscow 123182, Russia*

[2]*Department of Physics and Astronomy, University of British Columbia, Vancouver, BC, Canada V6T 1Z1*

[3]*Bruker BioSpin AG, Industriestrasse 26, 8117 Fallanden, Switzerland*

[4]*Department of Physics, Texas Tech University, Lubbock, Texas 79409-1051, USA*

[5]*School of Physics and CSEC, The University of Edinburgh, Mayfield Road, Edinburgh EH9 3JZ, UK*

**In strongly correlated materials, cooperative behaviour of the electrons causes a variety of quantum ordered states that may, in some cases, coexist[1-3]. It has long been believed, however, that such coexistence among ferromagnetic ordering, superconductivity and heavy-fermion behaviour is impossible, as the first supports parallel spin alignment while the conventional understanding of the latter two phenomena assumes spin-singlet or antiparallel spins. This understanding has recently been challenged by an increasing number of observations in uranium intermetallic systems (UGe$_2$, URhGe, UIr and UCoGe)[4-7] in which superconductivity is found within a ferromagnetic state and, more fundamentally, both ordering phenomena are exhibited by the same set of comparatively heavy 5$f$ electrons. Since the coexistence of superconductivity and ferromagnetism is at odds with the standard theory of phonon-mediated spin-singlet superconductivity, it requires an alternative pairing mechanism, in which electrons are bound into spin-triplet pairs by spin fluctuations[8,9]. Within the heavy-fermion scenario, this alternative mechanism necessarily assumes that the magnetism has a band character and that said band forms from heavy quasiparticles composed of $f$ electrons. This band is expected to be responsible for all three remarkable phenomena — heavy-fermion behaviour, ferromagnetism and superconductivity — although its nature and the nature of those heavy quasiparticles still remains unclear. Here we report spectroscopic evidence (from high-field muon spin rotation measurements) for the formation in UGe$_2$ of subnanometer-sized spin polarons whose dynamics we follow into the paramagnetic and ferromagnetic phases. These spin polarons behave as heavy carriers and thus may serve as heavy quasiparticles made of 5$f$ electrons; once coherence is established, they form a narrow spin-polaron band which thus provides a natural reconciliation of itinerant ferromagnetism with spin-triplet superconductivity and heavy-fermion behaviour.**

Within the BCS theory of superconductivity (SC), it became clear long ago[10] that pairing of electrons in the spin-singlet state is effectively destroyed by an exchange mechanism arising from strong Coulomb interactions between the valence electrons. In a ferromagnetically (FM) ordered state, this exchange interaction tends to align the spins of electrons within a Cooper pair in parallel, thereby effectively preventing the pairing. Likewise, within the standard heavy-fermion (HF) approach, the Kondo effect quenches the on-site magnetic moment below the Kondo temperature ($T_K$) by spin fluctuations (flips caused by interactions between the conduction electrons and localized *f*-electrons of the magnetic ions), thereby destroying pairs. However, such antagonism can be effectively avoided if the HF behaviour does not involve Kondo scattering[11,12].

Although these three phenomena — magnetism, superconductivity and heavy-fermion behaviour — have been considered in the past to be mutually antagonistic, the following findings clearly establish their possible coexistence (for a recent review, see Reference 3). A distinctive class of *f*-electron Ce- or U-based systems[2,3,13,14] convincingly shows that HF behaviour may coexist with SC. Most importantly, the SC pairing occurs *among the heavy quasiparticles* rather than within a band of light electrons, as was first demonstrated in $CeCu_2Si_2$[15] and uranium HF compounds[16,17]. In these materials, the scale of the specific-heat anomaly at the superconducting transition temperature $T_{SC}$ clearly demonstrates not only a large density of states associated with itinerant quasiparticles but, more fundamentally, that the SC energy gap opens up *within* the band of heavy quasiparticles. Since the strong mass enhancement in HF systems goes hand-in-hand with a dramatic renormalization of the heavy quasiparticle bandwidth, the characteristic quasiparticle velocity (Fermi velocity) is reduced by several orders of magnitude. This circumstance violates the fundamental requirement for phonon-mediated pairing, specifically that the sound (phonon subsystem) velocity is much lower than the quasiparticle velocity — the so-called time-delayed charge-charge interaction[3,18] — which then fails to suppress the direct Coulomb repulsion, making the phonon coupling mechanism ineffective. This fact alone indicates that the attractive interaction between the quasiparticles is probably *not* provided by the electron-phonon interaction as in ordinary BCS superconductors, but rather calls for an alternative mechanism which is offered by various spin-fluctuation models[1,2,8,9,18,19] of magnetically mediated SC.

Furthermore, in these HF materials SC may coexist and couple with magnetism. In fact, many *f*-electron HF systems exhibit SC deep within magnetically ordered states, suggesting that magnetism may *promote* rather than destroy the superconductivity[3]. It is remarkable that in this class of materials the same set of heavy quasiparticles apparently supports both the magnetism and superconductivity[13,14]. In particular, both SC and HF behaviour are suppressed in $CeCu_2Si_2$ when magnetic $Ce^{3+}$ ($4f^1$) ions are replaced by nonmagnetic $La^{3+}$ ($4f^0$) ions[15]. Coexistence of superconductivity and magnetism thus constitutes a clear distinction from other classes of SC magnetic materials[20-23] where fundamentally different electron subsystems are responsible for the two phenomena.

Although there is a growing consensus that HF *f*-electron materials do exhibit unconventional forms of magnetically mediated SC, the majority of systems studied so far support unconventional spin-singlet pairing mediated by antiferromagnetic (AFM) fluctuations[3,13-19], as opposed to ferromagnetic coupling, which is expected to adopt a spin-triplet configuration. However, an increasing number of observations in uranium HF compounds clearly demonstrate coexistence of SC and ferromagnetism. So far, this list includes $UGe_2$[4], $URhGe$[5], $UIr$[6] and $UCoGe$[7]. In all of these materials, the SC state is detected *within* the FM ordered state, at either ambient or elevated pressures.

In the theory of magnetically mediated superconductivity, it is important that the ferromagnetism itself is of *itinerant* character, similar to that in the canonical *d*-electron ferromagnets Fe, Co or Ni. In the majority of the HF compounds, the 5*f* orbitals are more localized, due to dominance of the strong intra-atomic Coulomb repulsion energy over corresponding bandwidths. However, in this group of compounds the 5*f* electrons seem to exhibit rather itinerant behaviour as a result of hybridization (mixing) with conduction band states[4,24,25]. This point requires somewhat deeper consideration.

Conventionally, electrons in solids are classified as either itinerant or localized. In strongly correlated electron systems, specifically HF systems, such a clear distinction is often obscured, since signatures of both pictures appear[13,14]. Here the strong Coulomb repulsion suppresses charge fluctuations at each site, leaving only spin and orbital degrees of freedom of localized states. These localized states interact with conduction electrons, and thereby affect one another. Heavy fermions are typically described by the Anderson-Kondo lattice models of coupled itinerant and localized electrons originating from different orbitals, in particular using a two-fluid description[26]. The possibility that the same electrons might simultaneously exhibit both localized and itinerant characteristics due to strong Coulomb interactions has developed into a duality problem[27] which, in fact, speaks to the heart of the debate concerning the nature of *f* electrons in condensed matter systems — are they localized, itinerant or of a dual nature (partially localized and partially itinerant)? In U- and Pu-based HF materials, a segregation has been proposed in which some of the 5*f* electrons are localized while the rest are itinerant. It has been suggested that two of uranium's three 5*f* electrons are localized close to the ionic core to produce AFM order, while the remaining *f* electron is delocalized to ensure SC in $UPt_3$[28] and $UPd_2Al_3$[29,30]. Similarly, four of the five plutonium *f* electrons are suggested to be localized and one to be itinerant in order to account for both magnetism and SC in $PuCoGa_5$[31]. Although such a duality model offers a transparent mechanism for producing quasiparticle mass enhancement — the exchange interaction between the itinerant and localized *f* electrons — it fails to receive support from de Haas van Alphen (dHvA) experiments[32]. Moreover, the huge amount of condensation entropy (on the order of the spin entropy) released at the SC transition clearly indicates that spins of *all* local moments participate in the formation of the SC order parameter[16]. In general, it is hard to find an explanation within the existing duality models of how competition between intra-atomic Coulomb interactions and anisotropic hybridization of *f* electrons (both on the order of eV) can differentiate between indistinguishable intra-atomic electrons and result in a ground state of coexisting magnetism and SC which is controlled by *f* electrons on an energy scale of 1 meV[33]. The situation is even more confusing if a *single f* electron, such as in cerium, were to display localized and itinerant character *simultaneously*, as is required within the duality model(s) in, for instance, $CeCu_2Si_2$[15], $CeIrIn_5$[33] and other Ce-based HF superconductors, to account for coexisting magnetism and SC. The electronic duality proposed for many different HF systems exhibiting simultaneous SC and magnetism would require *the same f-electron* to display both localized and itinerant nature *simultaneously*. This fundamental problem requires a new conceptual framework in which an appropriate description of strong electronic correlations with theoretical access to low energy scales must be a key ingredient[33].

Here we propose a specific concept that may supply the necessary requirement of simultaneously itinerant and localized electrons: formation of a spin-polaron band in which quasiparticle excitations of a low energy scale (several meV) around the Fermi energy ($E_F$) are responsible for HF behaviour, SC and magnetism.

The 5$f$ electron duality problem is still debated in UGe$_2$, which is proposed to be viewed as a two-subset electronic system, where some of the 5$f$ electrons are localized and responsible for the ferromagnetic moment and huge magnetocrystalline anisotropy, while the remaining 5$f$ electrons are itinerant and responsible for unconventional SC[34]. The nearest distance between U atoms in UGe$_2$, 0.39 nm, far exceeds the Hill limit, so without hybridizing with conduction electrons, the 5$f$ electrons will be localized[3,35]. The Curie-Weiss-like temperature dependence of the susceptibility, the strong anisotropy, the large orbital moment of the U atom, the lack of induced magnetization at Ge sites and several other features[3,35,36] all indicate a local character of 5$f$ electrons in UGe$_2$. On the other hand, there is strong experimental evidence for their itinerant nature. In particular, UGe$_2$ exhibits rather good agreement between specific heat[35,37] and de Haas van Alphen[38-40] results, which both attest to the itinerant character of the heavy quasiparticles; these data yield an effective mass $m^* \sim 10\text{-}25 m_0$ at ambient pressure ($m_0$ is the free electron mass), similar to $m^*$ in the itinerant-electron 3$d$ ferromagnet MnSi. Furthermore, itinerant behaviour of the 5$f$ electrons in UGe$_2$ is suggested by Hall-effect[41] and muon spin relaxation[42] measurements and is also frequently discussed for other uranium HF compounds[28,43]. The band picture is also consistent with spontaneous magnetization with non-integer Bohr magneton ($\mu_B$) values per atom — 1.48$\mu_B$ in UGe$_2$[4,35], 0.42$\mu_B$ in URhGe[5], 0.5$\mu_B$ in UIr[6] and 0.03$\mu_B$ in UCoGe[7] — which is significantly smaller than the Curie-Weiss moment (2.7 $\mu_B$/ion in UGe$_2$[35]) detected in the paramagnetic state above $T_{Curie}$ — again similar to 3$d$ band magnetism in MnSi[44].

The magnetism in itinerant FM systems originates from exchange splitting of the band states rather than from strongly localized electrons. The mass enhancement comes from an extreme renormalization of band widths ($\Delta$) — from $\Delta_o \sim 1\text{-}10$ eV, typical of conventional metals, to renormalized $\Delta \sim 0.001\text{-}0.01$ eV in the canonical HF systems in which the $f$-electron degrees of freedom are typically modelled as hybridized with conduction band states to yield heavy itinerant quasiparticles that form extremely narrow spin-split bands in the vicinity of the Fermi surface[24]. In the FM state, within such a spin-polarized band the heavy quasiparticles would have difficulty forming ordinary spin-singlet Cooper pairs but may instead select unconventional configurations[4,35] that involve non-zero spin and angular momentum states analogous to those in $^3$He, in which a spin-fluctuation mechanism is responsible for the formation of spin-triplet pairs[1]. Thus we arrive at a familiar picture of a band of delocalized states, albeit rather heavy, that would be responsible for FM and SC. However, the nature of this band and the nature of said delocalized states (heavy quasiparticles) both have yet to be determined.

The standard theory for the formation of the extremely narrow high mass bands characteristic of heavy fermion metals, such as UGe$_2$, starts from a set of strongly localized *f*-electrons. The appearance of a new (low) energy scale in this approach results from hybridization with the delocalized conduction states and strong correlations within the *f*-shells. A different approach, that we wish to develop here, starts with a delocalized band carrier whose transport depends upon the strength of its coupling with excitations of the medium. This is similar to the case of a lattice polaron[45] (LP), where in the limit of strong coupling an electron accompanied by lattice modes (displacements of ions) forms a quasiparticle in which a local distortion of the crystal structure follows the charge carrier adiabatically and whose bandwidth $\Delta_{LP}$ is reduced by up to 4 orders of magnitude relative to that of ordinary electrons in conventional metals[46]. At low temperature, as long as the polaron is still much lighter than the atoms composing the medium, the charge is then delocalized within the LP band[47,48]. A remarkable collapse of $\Delta_{LP}$ at higher temperature marks a crossover from coherent band dynamics to incoherent hopping of localized states, analogous to the so-called "dynamic destruction of the band"[49,50] for the tunnelling dynamics of heavy particles, such as protons, isotopic defects, or muons and muonium[51]. In close analogy, the exchange interaction (*J*) between a free carrier and local spins can cause electron localization into a FM "droplet" on the scale of the lattice spacing in a paramagnetic (PM) or AFM "sea"[47]. This charge carrier, accompanied by reorientations of local spins, forms a *spin polaron*[46,47,52] (SP) — a "giant spin molecule"[53] — with a large composite spin (*S*). As in the case of the LP, formation of a SP profoundly renormalizes the "bare electron" band into an extremely narrow ($\Delta_{SP} \sim 0.001$-$0.1$ eV) spin-polaron band, which will favour coherent SP band dynamics at low temperature as long as spin fluctuations are suppressed[46,47].

Within this conceptual framework the *composite quasiparticle* (spin polaron) consists of the light, initially delocalized bare carrier (say, an *s*- or *p*-electron) from the initially non-renormalized band ($\Delta_o$) which undergoes localization within the SP, and a set of local *f*-spins confined within the scale of the electron's wavefunction. Such a composite quasiparticle is free to propagate within the renormalized, though coherent, SP band ($\Delta_{SP}$) *via* coherent reorientations of local *f*-spins[47] in essentially the same way as another composite quasiparticle — the famous Landau lattice polaron[45] (a bare electron plus lattice displacements or phonons). Once coherence among such spin polarons is established within the SP band, a remarkable result is that heavy *f*-like quasiparticles (SP) become part of the Fermi surface. We specifically note that within this framework the heavy quasiparticle is not an *f*-electron but the composite quasiparticle, a SP, formed as a result of *s*(*p*)-*f* interaction.

Regardless of these heavy quasiparticles' origin, an issue of a fundamental importance is that, being part of the Fermi surface, they should obey the principles of the Landau Fermi liquid — in particular, the counting rule or Luttinger's theorem[54], which states that, in non-interacting electron band theory, the volume of the Fermi surface ($V_{FS}$) counts the number of conduction electrons ($n_e$). For interacting systems, this rule changes[55-57] to manifest a remarkable result that the local spin states ($n_{spins}$) are also included into the sum: $2V_{FS}/(2\pi)^3 = n_e + n_{spins}$. Therefore, even though *f*-electrons are localized as magnetic moments at high temperature, they contribute to the Fermi surface volume in the heavy Fermi liquid[57]. This fundamental point of the heavy fermion physics is often discussed in terms of the transformation from a "small" Fermi surface containing only conduction electrons to a "large" Fermi surface which includes both conduction electrons and local spins[56,57]. Related issues arise in the context of a quantum critical point separating the heavy-fermion paramagnet from the local moment magnet, at which point the Fermi surface contracts from a large to a small volume[56,57]. We argue that formation of a spin-polaron band may not only give a natural description of how *f*-spins, being local moments, nevertheless acquire itinerancy (the duality problem), but may also pave a way to understanding of how Luttinger's theorem works in strongly correlated materials. Such an approach finds its experimental confirmation in quantum oscillatory studies, Hall measurements or optical experiments, all of which indicate that the Fermi surface reconstructs to include *f*-spins at low temperature or contracts when composite quasiparticles disintegrate at a quantum critical point[57].

In 5*f* UGe$_2$, formation of an extremely narrow spin-polaron band might be expected to occur in a manner very similar to the SP band formation in 4*f* systems[46], since the *p(d)-f* exchange constant in UGe$_2$ is rather large ($J = 0.44$ eV)[25], comparable to that found in the 4*f* Eu and Sm chalcogenides[46] where $J \approx 0.5$ eV. Dramatic renormalization of such a SP band is expected to be associated with a change in the coherence properties and will go hand-in-hand with a significant increase of the electron effective mass, which might allow for the application of general concepts developed for the coherent-to-incoherent crossover of the heavy particle tunnelling dynamics[49,51]. For such particles, one encounters different behaviour in metals and insulators due to the essentially different spectral properties of the environments: incoherent dynamics take over at high temperature ($k_B T \gg \Delta_{SP}$) in insulators while suppression of the coherence in a metal is expected at temperatures as low as $k_B T \sim \Delta_{SP}$[49,51]. Thus at high temperature, the SP dynamics are characterized by incoherent hopping (diffusive) transport and the band picture, with its Bloch states, breaks down. However, in the low-temperature limit, the band picture manifests itself in coherent transport of heavy quasiparticles — the spin polarons. Such thermal destruction of spin-polaron bands has recently been reported in correlated 3*d*-electron semiconductors[58] and 5*d*-electron metallic host[59]. Thus, a description involving SP allows one to arrive at results qualitatively similar to those produced by the standard approach to heavy fermion systems, albeit without the involvement of Kondo screening.

Here we present spectroscopic evidence, obtained by positive muon spin rotation ($\mu^+$SR), for spin polarons in UGe$_2$, confined within $R = 0.25(1)$ nm, with a high spin of $S = 4.3 \pm 0.3$. At low temperature, SP tend to form a narrow spin-polaron band in the vicinity of $E_F$, profoundly modifying the magnetic, transport, optical and thermodynamic properties of the host.

To develop a physical picture we consider a charge carrier (electron which has a strong exchange interaction ($J$) with surrounding magnetic ions[52,53]. In a magnetic system, the electron's energy depends strongly on the magnetization, with the minimum energy achieved by FM ordering[46]. As long as the direct coupling between ions is comparatively weak, the indirect coupling of magnetic ions mediated by this shared electron can cause local FM ordering accompanied by strong electron localization[46,47,52,53,60]. Such localization inevitably involves a significant increase of the electron's kinetic energy — which can, however, be compensated by the corresponding energy reduction associated with the local FM ordering of the ions mediated by the aforementioned electron. Thus, the increase of electron kinetic energy, due to confinement, may be compensated by the difference in exchange energy between the final locally saturated FM region and the initial paramagnetic, AFM or even non-saturated FM state. Therefore, the electron tends to establish and support this local ordering, thereby forming a FM "droplet" over the extent of its wave function (typically on the order of the lattice spacing, $a$)[46,47,52,53]. This charge carrier, accompanied by reorientations of local spins to form its immediate FM environment, together behave as a single quasiparticle — a spin polaron — with a composite spin ($S$)[46,47,60]. In magnetic semiconductors (MS), in the process of electron localization at a donor impurity (an implanted muon in our case), the discussed increase in the kinetic energy, assisted by the entropy change ($\Delta S$) due to ordering within the SP, is compensated by the combined efforts of the exchange interaction and the Coulomb interaction with the corresponding donor. The net change in the free energy

$$\Delta F = \hbar^2/2m^*R^2 - Ja^3/R^3 - e^2/\varepsilon R + T\Delta S \qquad (1)$$

has a minimum as a function of $R$, which represents the radius of electron confinement (or equivalently, the extent of its wavefunction). $\Delta F$ decreases with decreasing $R$ until $R \ll a$, at which point the electron wavefunction no longer overlaps even the nearest ions, and the exchange term vanishes[46]. The exchange energy [second term in Eq. (1)] is optimized by maximizing the SP electron's net overlap with the $f$-shells of nearby ions. The Coulomb interaction [third term in Eq. (1)], important in MS[58,60-65], is effectively screened in metals[59,66]. Once the Coulomb interaction can be neglected, one immediately arrives at the initial idea of de Gennes on carrier confinement within free delocalized SP[52]. In UGe$_2$, the high electron concentration ($\sim 10^{22}$cm$^{-3}$)[41] ensures that the screening length of the Coulomb term is much smaller than $R$. At low temperature, the entropy term $T\Delta S$ is also small. Each of the remaining two terms in Eq. (1), namely the electron's kinetic energy and the exchange energy, is on the order of an eV and not only far exceeds any other energy scale in the problem but also reduces the length scale for electron confinement to within less than one unit cell[59,66].

Thus far, extensive studies of the formation and dynamics of spin polarons in magnetic semiconductors, magnetoresistive perovskites and related compounds[46,61,67] have been restricted to a quite narrow temperature range close to a magnetic phase transition (large polaron region). Different macroscopic techniques (*e.g.*, SQUID magnetometry or magnetotransport) are spatially averaged, providing little information on possible spatial inhomogeneities. The spatial resolution of magneto-optical measurements[68], photoluminescence[69] or soft X-ray magnetic circular dichroism[70] is severely restricted by wavelengths on the order of 10 nm. Studies of spin polarons in MS by microscopic techniques like NMR[71] or Raman scattering[72] are restricted to the close vicinity of a transition by limitations in their sensitivity that makes it essentially impossible to detect a spin polaron as it shrinks towards the sub-nm scale. Although the significantly better spatial resolution of the small-angle neutron scattering technique made it possible to detect spin polarons of about 1.2 nm size in magnetoresistive perovskites just above the transition temperature, this technique is still limited to a narrow temperature range by the neutron wavelength of about 0.5 nm, which made it impossible to detect smaller spin polarons[73]. In general, the limited sensitivity and/or spatial resolution of many different techniques precludes detection of a SP of subnanometer size.

Here the unique sensitivity of polarized positive muons as a local magnetic probe makes muon spin rotation and relaxation ($\mu^+$SR)[74-76] ideally suited for mapping the magnetic state on the atomic (sub-nm) scale. As the spin polaron expands toward a 10 nm scale, this sensitivity advantage is lost, making $\mu^+$SR complementary to a variety of other techniques mentioned above. This approach has already been applied to studies of SP in different materials ranging from insulators (including AFM[77]) to itinerant ferromagnets[66] analogous to earlier studies of nonmagnetic semiconductors[78], which revealed the details of electron capture to form the muonium (Mu ≡ $\mu^+e^-$) atom (a light analogue of the H atom)[79-82]. Assorted SP have recently been detected in 4*f* and 3*d* magnetic semiconductors[58,60,62-65,83], in the 5*d* and 3*d* correlated metals[59,66] and in a quasi-1D AFM insulator[77] *via* $\mu^+$SR spectroscopy.

Single crystals of UGe$_2$ for the current studies were grown by the Czochralski technique under purified Ar atmosphere with a water-cooled crucible and radio-frequency heating. Single crystals were oriented using a white beam X-ray backscattering Laue method, sparkcut and etched to remove the oxidised surface. The electrical resistivity was measured by the conventional four-probe method between 1.8 and 300 K in zero magnetic field. Small crystals cut off the single crystals used in $\mu^+$SR experiments showed a residual resistivity ratio RRR=62 and $T_{\text{curie}}$= 52.6 K.

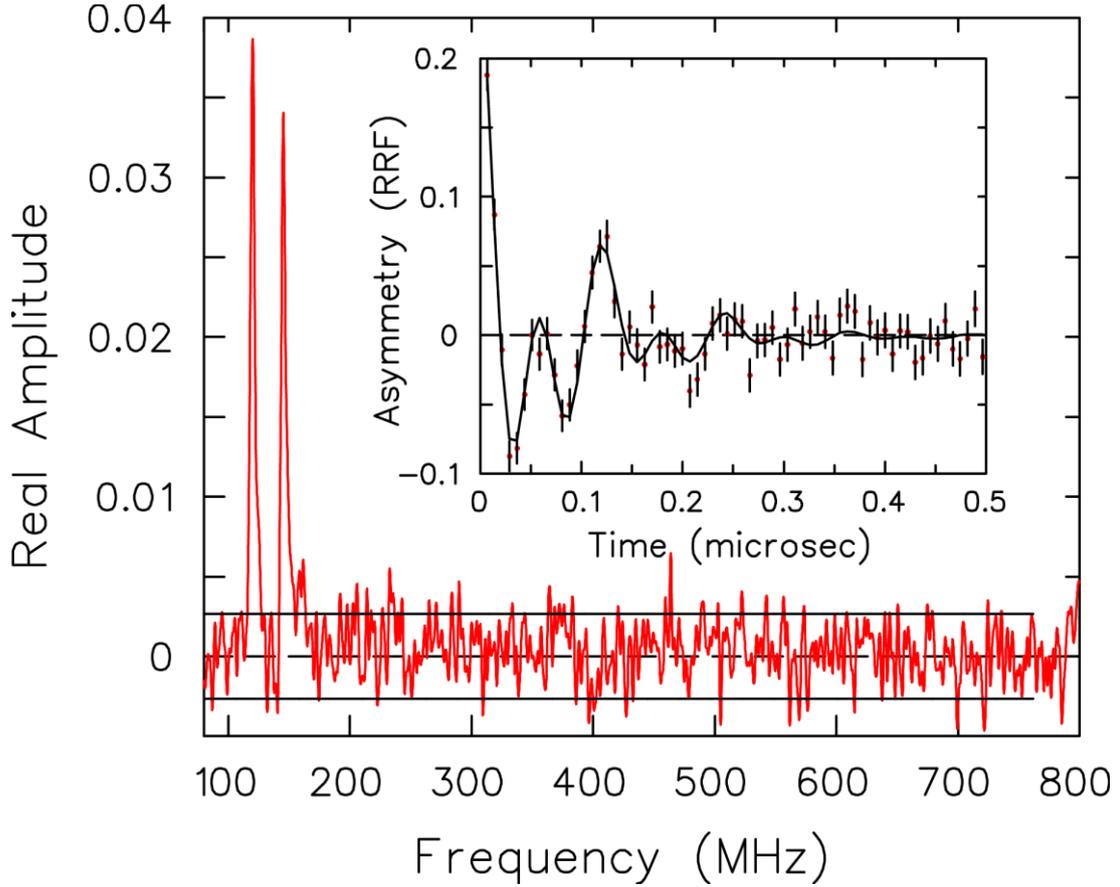

**Figure 1** Frequency spectrum of muon spin precession in UGe$_2$ in a transverse magnetic field of $H = 1$ T at $T = 40$ K. Only the real part is shown, as including the imaginary part artificially broadens the overall lineshape[75]. Inset: same spectrum in the time domain transformed into a rotating reference frame[75] at 135.53333 MHz. The two-frequency precession pattern characteristic of a localized electron hyperfine-coupled to a muon is clearly apparent in both domains.

Time-differential $\mu^+$SR experiments, using 100% spin-polarized positive muons implanted into these samples, were carried out on the M15 surface muon channel at TRIUMF using the *HiTime* spectrometer. At high temperature, Fourier transforms of the $\mu^+$SR time spectra in a magnetic field (***B***) transverse to the initial muon spin polarization direction and parallel to the easy magnetization direction (***a*** axis) of the single crystal of UGe$_2$ exhibit a single peak at the muon frequency $\nu_\mu = \gamma_\mu B/2\pi$ (where $\gamma_\mu/2\pi = 135.53879$ MHz/T is the muon gyromagnetic ratio). However, below $T \approx 100$ K the $\mu^+$SR spectra change abruptly to reveal two peaks (Figure 1) — a characteristic SP doublet similar to that in another itinerant FM system, MnSi[66]. The evolution of SP signals with temperature is presented in Figure 2. These two peaks are also shifted to lower frequencies (although to a lesser amount than in MnSi) relative to the single peak (not shown in Figure 1) detected in a reference sample (CaCO$_3$), which occurs at the bare muon frequency. The smaller shift detected in $UGe_2$ is consistent with its lower magnetization compared to that in MnSi, and similar to the corresponding shifts in magnetic semiconductors, which empirically scale with the magnetization[60,62]. Utilizing the same experimental setup as used for measurements with MnSi[66] ensured that there is essentially no background in our $UGe_2$ signals — the muons that miss the sample and stop in $CaCO_3$ are detected in a different combination of counters and routed to independent histograms which form the reference signal. This reference frequency does not depend on temperature, since $CaCO_3$ is nonmagnetic, and therefore provides an independent monitor of the applied magnetic field.

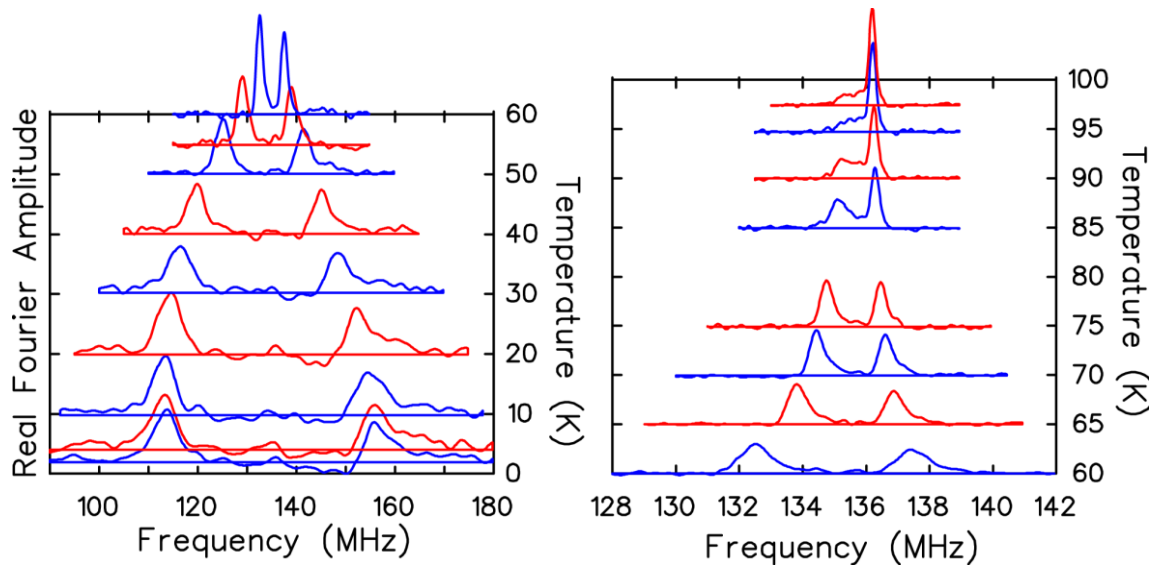

**Figure 2** Fourier transforms of the muon spin precession signal in $UGe_2$ in a transverse external magnetic field of 1 T at different temperatures. The characteristic SP lines appear below about 100 K and persist through the FM transition ($T_c$ = 52.6 K) down to the lowest measured temperature. Note that the frequency scale changes by a factor of 0.156 between low and high $T$, reflecting the dramatic reduction of the splitting at high $T$. Also note the qualitative change in the line shapes around 80 K.

There are, however, several important differences in the Fourier spectra observed in MnSi and $UGe_2$. First, in MnSi the characteristic two-frequency spectra persist up to room temperature, whereas the doublet pattern is detected only below about 100 K in $UGe_2$. Second, the amplitudes of the two lines are temperature independent and almost equal in MnSi over the entire temperature range, whereas in $UGe_2$ they exhibit a remarkable temperature dependence: they are temperature independent below about 80 K, but between 80 K and 100 K the $\mu^+$SR doublet spectra exhibit a strong temperature dependence with increasingly different amplitudes of the two lines (Figure 3). We discuss these two essential points below.

Observation of two peaks in the Fourier spectra prompted the authors of Ref. 84 to suggest two magnetically inequivalent sites occupied by the positive muon in $UGe_2$. Although such an approach constitutes the conventional assignment of multiple signals in a magnetically ordered state, fast spin fluctuations make this interpretation irrelevant in the paramagnetic phase[59,60,62-66], while possible Knight shifts from the conduction electrons are typically at least 2-3 orders of magnitude smaller than the characteristic splittings detected in this experiment[74]. Moreover, the two peaks do not follow the temperature dependence of the magnetization, which clearly indicates that the muon does not stay "bare" and act as local magnetometer. Instead, while one peak goes up in frequency the other goes down as temperature varies (Figure 4) — the fingerprint of a muon-electron bound state[75]. Furthermore, a qualitative change in the amplitude ratios within the doublet (Figures 2 and 3) is obviously incompatible with the two-site interpretation, which assumes a constant muon site occupation ratio. Finally, in the PM state the Knight shifts are expected to be linear with magnetic field, which contradicts the experiment (Figure 5). This line of argumentation is similar to that presented for another HF superconductor, $UBe_{13}$, which also exhibits SP[85].

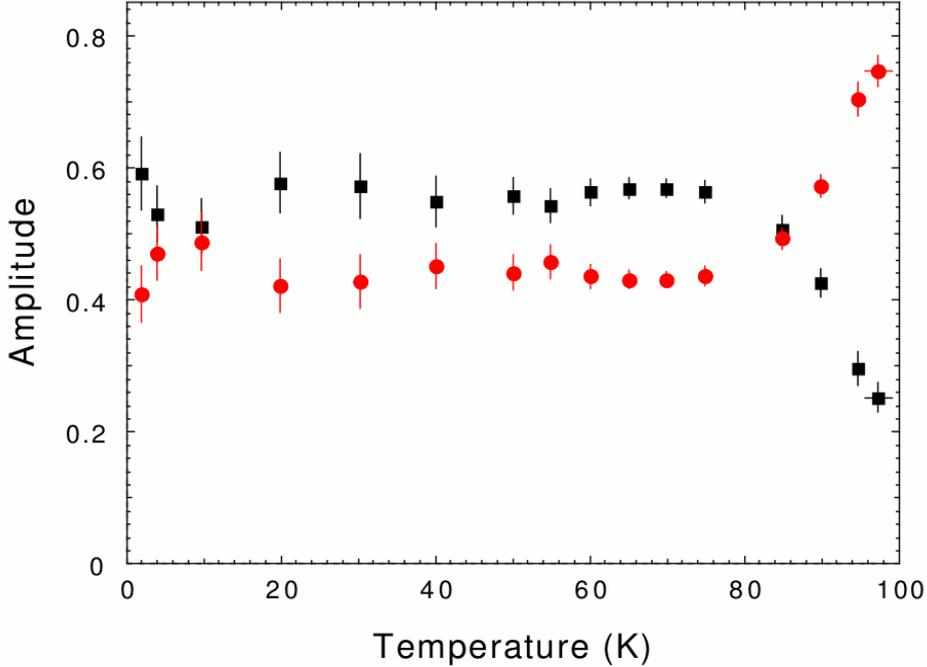

**Figure 3** Temperature dependence of the amplitudes of muon spin precession signals (spectral weight or population) in $UGe_2$ in a transverse external magnetic field of 1 T. Red circles and black squares show the evolution of the signal with the higher and lower frequency, respectively, around 100 K. Note the qualitative change around 80 K.

Instead, we argue that the two lines shown in Figure 1 and Figure 2 constitute the characteristic signature of a coupled muon-electron spin system in high magnetic field[59,60,62-66,74,75,78,86]. The solution of the Breit-Rabi Hamiltonian which governs a muon-electron spin system yields 4 eigenvalues (due to 4 possible combinations of spins) corresponding to 4 energy levels with different allowed transitions[78]. In high magnetic field, the two allowed transitions correspond to the two allowed muon spin-flip transitions between states with fixed electron spin orientation; the frequency splitting between these two transitions is determined by the muon-electron hyperfine interaction $A$[74,75,78]. Moreover, we argue that the observed bound state is a spin polaron. In a PM or metallic (or both) environment, the strong pair exchange interaction of the bound electron with itinerant spins (spin exchange[74,78]) would result in rapid spin fluctuations of this electron, averaging the hyperfine interaction to zero — which, in turn, would result in a collapse of the doublet into a single line at $v_\mu$ (see Ref. 86 for details), if the local FM ordering mediated by this electron did not hold the electron's spin orientation "locked"[59,60,62-66,85,86]. In metals, however, even the protective local FM environment of a SP does not ensure observation of the doublet unless the SP spin ($S$) is decoupled from its magnetic environment[59,66,86]. Such decoupling is possible in high $B$ when the Zeeman energy of $S$ exceeds an exchange interaction ($I$) between local spins[60,85]. This is the case in magnetic insulators where the SP doublet is detected up to very high temperature[60,62-65]. In metals, RKKY interactions make $I$ much stronger, so that decoupling would require a very high magnetic field[85] that is inaccessible in the current experiment. In UGe$_2$, above about 100 K, spin exchange with the magnetic environment effectively averages the muon–electron hyperfine interaction, causing the collapse of the doublet into a single peak, as observed, discussed and presented here in Figures 2 and 4.

Such a collapse by no means signifies that the SP does not form in UGe$_2$ above 100 K; we just do not see its fingerprint, which is a characteristic doublet[85,86]. The abrupt appearance of a SP doublet below about 100 K is possible due to another effective decoupling mechanism — the opening of a spin gap due to crystal field splitting of the U ion's spin excitation, characteristic of U compounds[34]. Crystal field splitting eliminates low-lying spin excitation from the spectrum, making spin exchange of the SP spin ($S$) with its magnetic environment rather ineffective. A detailed evaluation of the crystal field splitting shows a spin gap of about 120 K between the almost degenerate ground state and the first excited state[34]. This circumstance makes it possible to detect the characteristic SP doublet below about 100 K in UGe$_2$, much like that in another heavy-fermion compound, UBe$_{13}$, that exhibits a spin gap of about 180 K[85]. Likewise, optical conductivity exhibits a dramatic reduction of the spin scattering rate below the characteristic energy of about 120 K[87]. By contrast, very strong crystal field splitting, characteristic of 3$d$ systems, effectively prevents spin exchange in MnSi, enabling detection of the SP doublet all the way up to room temperature[66] in that 3$d$-electron counterpart of UGe$_2$. A spin gap opening in the U ion spin excitation spectrum, although a necessary condition, does not prevent spin exchange between SP and conduction electrons, which would cause the doublet to collapse. However, the dominance of the orbital moment of U ions in UGe$_2$[88] makes spin exchange with conduction electrons (which have zero orbital moment) again rather ineffective due to the orbital moment conservation law. Thus, the conditions for observation of SP in a metallic or PM (or both) environment are rather specific: one must eliminate any possible spin exchange mechanism in order to prevent the muon-electron hyperfine coupling from averaging to zero. However, the mere observation of the characteristic doublet in the $\mu^+$SR spectra of a metal constitutes strong evidence for SP formation[59,66,86]. The evolution of the two signals within the doublet as a function of temperature (Figures 2 and 4) and magnetic field (Figure 5) is consistent with that of the muon-electron bound state when the electron spin is locked to the SP spin, providing strong support for this picture.

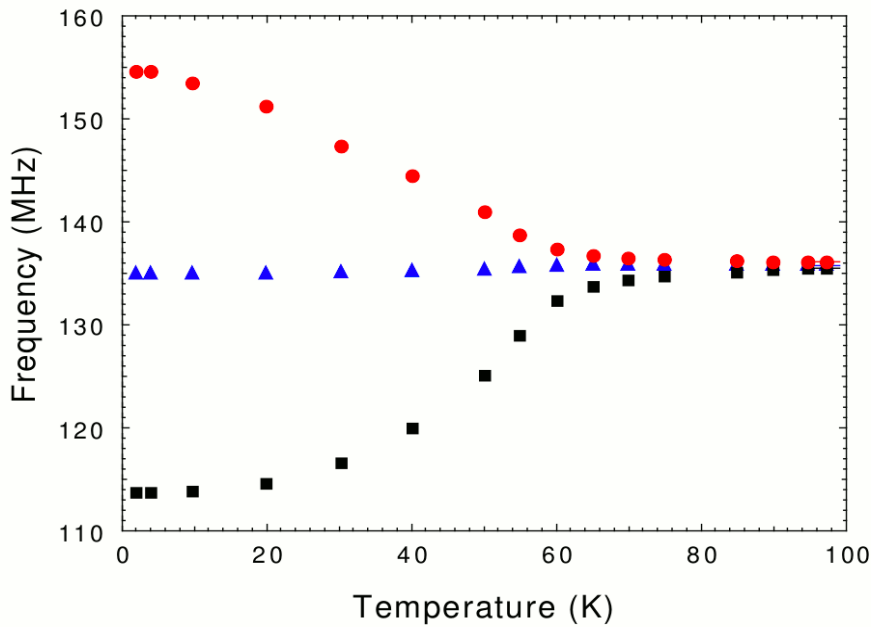

**Figure 4** Temperature dependences of muon spin precession signals frequencies in UGe$_2$ in a transverse external magnetic field of 1 T. The SP lines appear below about 100 K where the doublet is composed of a higher (red circles) and lower (black squares) frequency signal. The signal (blue triangles) from [nonmagnetic] CaCO$_2$, mounted directly behind the sample, provides a direct measure of the magnetic field at the sample *via* the measured frequency.

Temperature and magnetic field dependencies of the SP signal frequency splitting (within the doublet), $\Delta v$, provide information on the characteristic size (the localization radius, $R$, for the electron confinement) through the hyperfine coupling ($A$) and determine the composite spin ($S$) of the SP[60,86]. Within a mean field approximation[89], $\Delta v$ is proportional to a Brillouin function. For $g \mu_B B \ll k_B T$, $\Delta v$ is a linear function of both $B$ and $1/(T-T_c)$ [60,86]

$$\Delta v = A \left[ \frac{g \mu_B B}{3 k_B (T - T_c)} \right] (S+1) \qquad (2)$$

where $k_B$ is Boltzmann's constant and $T_c$ is the Curie temperature of UGe$_2$. At low $T$ and high $B$, however, equation (2) is no longer valid, as the composite spin ($S$) is fully polarized. Therefore, in a magnetic field high enough that the muon Zeeman energy exceeds the hyperfine coupling, $\Delta v$ saturates at the value of $A$[60,74,75,78]. In UGe$_2$, the splitting ($\Delta v$) saturates as a function of both inverse temperature (in a magnetic field $H = 1$ T) and $H$ (at $T = 5$ K) at the same value, $A = 41\pm2$ MHz (see Figure 5).

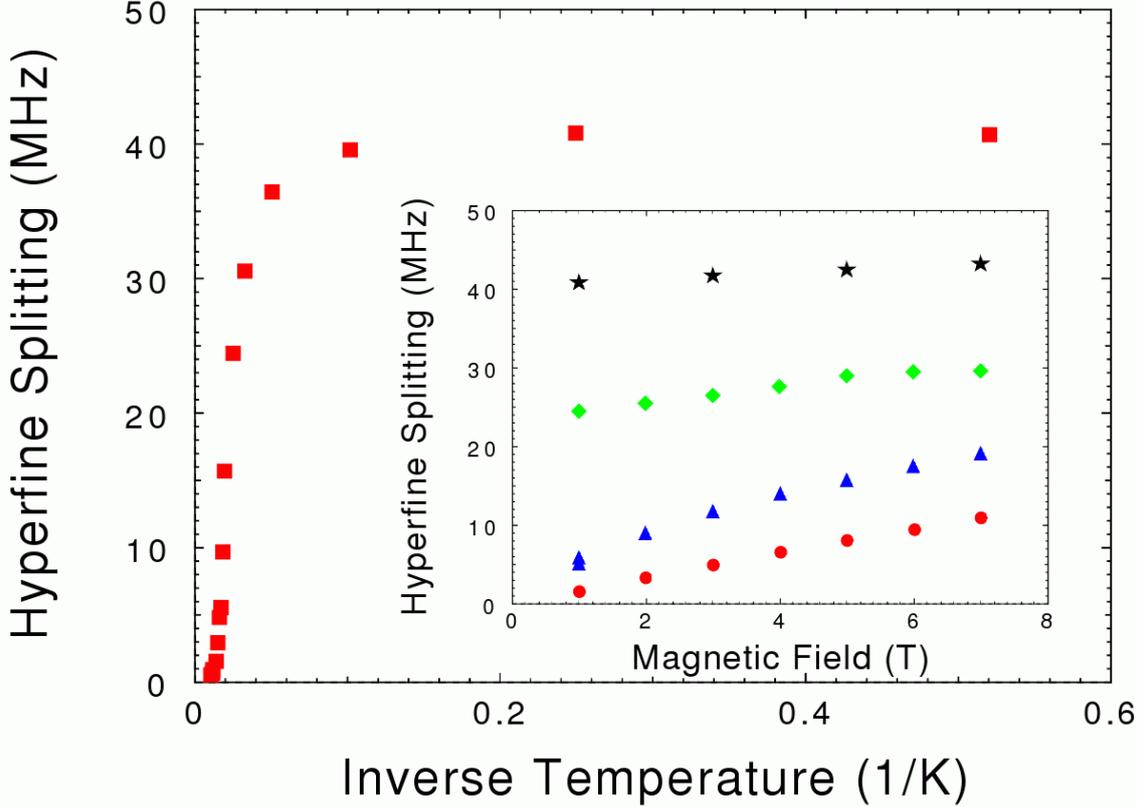

**Figure 5** Temperature dependence of the SP frequency splitting ($\Delta v$) in UGe$_2$ in a magnetic field of $H = 1$ T. At low temperature, the SP is fully polarized and the splitting saturates at the full strength of the $\mu^+e^-$ hyperfine coupling ($A$). Inset: magnetic field dependence of $\Delta v$ at $T = 75$ K (red circles), $T = 60$ K (blue triangles), $T = 40$ K (green diamonds) and $T = 5$ K (black stars). The saturation of both temperature and magnetic field dependences of $\Delta v$ are characteristic of hyperfine-coupled $\mu^+e^-$ spin systems. Both curves saturate at the same value of $A = 41\pm2$ MHz.

For our model of a SP captured by a muon, the SP electronic wave function is an extended hydrogen or muonium (Mu) atom $1s$ function, for which the value of $A$ scales as $A_{vac} (a_0/R)^3$, where $R$ is the characteristic Bohr radius (radius of the electron confinement) and $A_{vac} = 4463$ MHz is the hyperfine frequency of Mu in vacuum (for which $R = a_0 = 0.0531$ nm — the Bohr radius of Mu in vacuum)[75]. In UGe$_2$ at low temperature, the hyperfine coupling in the SP is about 100 times less than that for Mu in vacuum, implying that the radius of the electron confinement is $R \approx 0.25(1)$ nm. This is consistent with the muon being centred between two U atoms ($x$=0.5, $y$=0.5, $z$=0.5) giving a muon-U distance of 0.214 nm. As the $f$-orbital radius is 0.0527 nm, this muon position ensures the maximum overlap of the SP electron with U $f$-wavefunctions, resulting in a SP composed of one SP electron and the two nearest U ions. We can estimate the composite spin ($S$) for this SP. The magnetic moment for such a SP is determined by 2 fully polarized U ions, each having $\mu_\text{U} = 2.7\mu_\text{B}$[4], minus the SP electron's $1\mu_\text{B}$, since the SP electron spin is antiparallel to the composite SP spin at high temperature (see below). Meanwhile, $\mu_\text{SP} = 2\mu_\text{U} - 1\mu_\text{B} = g \cdot \mu_\text{B} \cdot [S(S+1)]^{0.5}$. From this relation, and accounting for the dominance of the orbital moment in UGe$_2$ (which causes $g = 0.8$)[88], we find $S \approx 5$. On the other hand, fitting Equation (2) to the $\mu^+$SR data in Figure 5 using $A = 41\pm2$ MHz and $g = 0.8$ yields $S = 4.3(3)$ and $T_\text{c} = 52(1)$ K, consistent with our estimate of $S$ from the SP radius and $T_\text{Curie}$ of our sample, respectively. At low temperature, the SP electron's spin flips to be parallel to the composite SP spin, causing the SP spin to increase to $S \approx 6$. For comparison, the SP detected in MnSi gives $A = 12\pm1$ MHz and $R \approx 0.4$ nm, corresponding to a rather large SP confinement radius within one unit cell of MnSi[66]. On the other hand, the size of the SP found in UGe$_2$ is, in fact, the same as that in UBe$_{13}$[86], while in the 5$d$ correlated metal Cd$_2$Re$_2$O$_7$ at low temperature the SP contracts to a rather compact 0.15 nm[59]. Apparently, in the orthorhombic structure of UGe$_2$ the SP prefers to localize within a uranium pair in order to maximize the gain in the exchange energy, which is necessary to ensure its localization.

Although the possibility of SP formation in AFM and PM states has long been anticipated[46-48,52,53] and experimentally established[58-73,77,85,86] in a great variety of materials by many different techniques, its existence would seem to be incompatible with the FM state. The exchange contribution to SP stabilization amounts to the *difference* between the PM disorder (or AFM order) of the host and the FM order within the SP; in a fully saturated FM state the exchange contribution to the localization would be negligible, as the lattice spins are already aligned. Such an increased alignment prevents SP formation in the FM state, as well as when a sufficiently high magnetic field is applied in the PM state in Heisenberg ferromagnets[60,62,63], where FM originates from localized electrons. In itinerant-electron ferromagnets, however, local spins are far from being saturated in the FM state, as evidenced by the rather low effective magnetic moments in UGe$_2$[4,35], URhGe[5], UIr[6] and UCoGe[7], as well as that in MnSi[66]. Therefore, electron localization into a SP that completes the spin alignment within the fully saturated SP core region is clearly possible within an itinerant-electron FM ordered phase[66]. Specifically, in UGe$_2$ the U atom spin saturation increases from its low temperature value $\mu_\text{U} = 1.5\mu_\text{B}$ to the fully polarized value $\mu_\text{U} = 2.7\mu_\text{B}$, making a huge difference in the exchange energy and ensuring electron localization *via* SP formation.

The exchange interaction governs spin polaron formation and dynamics in UGe$_2$, since the Coulomb interaction is effectively screened [see Equation (1)]. Therefore the role of the muon, which may be important for SP formation in MS[60,62-65] and insulators[77], is reduced to that of an "innocent bystander" microscopic magnetic probe in metals. We argue that once the host lattice is populated by free SP, one of them is captured by the muon, as in $Cd_2Re_2O_7$[59], $MnSi$[66] and $UBe_{13}$[85], to reveal the fingerprint of a bound muon-electron state — the characteristic SP doublet.

In $UGe_2$, the dynamics of such SP clearly shows a qualitative change around 80 K. At high temperature, the asymmetric doublet shape (Figure 2, right panel) indicates the nearly static character of the SP. The spectral weights (populations) of the spin-up and spin-down states, being determined by a thermal (Boltzmann) distribution, should be almost equal as the Zeeman splitting $\mu_{SP}H \ll k_BT$, which is apparently not the case (Figure 3). The temperature fails to equilibrate the amplitudes (populations) of the spin-up and spin-down states because the composite SP spin ($S$) is strongly decoupled from the magnetic environment. This decoupling occurs because spin exchange with both local spins and conduction electrons is rather ineffective, making it possible to detect the characteristic SP doublet in the first place (see discussion above). The difference in population of the two spin states is due to one of them being parallel to the magnetic field, resulting in a long-lived eigenstate, while the other state (antiparallel to the field) is short-lived or unpopulated[59]. The temperature dependence of the corresponding line widths (relaxation rates) clearly reflects this difference: at high temperature the line width of the stable eigenstate (higher frequency line) is much less than that of the other state (Figure 2 and Figure 6). Similar asymmetric distributions of the spectral weight between the two lines of the SP doublet are detected in magnetic semiconductors and insulators where the SP is found to be static[60,62-64,77].

A remarkable crossover occurs at about 80 K, where not only do the line widths become equal (Figure 6), but also the spectral weights (corresponding state populations) of the two lines within the doublet effectively equilibrate (Figure 3), clearly indicating the onset of the effective spin-exchange mechanism *within the SP system,* as other spin-exchange channels are rather ineffective (see above). This marks a crossover from the static SP to itinerant spin polaron behaviour. At high temperature, the characteristic length (in Angstroms) of the ferromagnetic fluctuations determined from neutron measurements[90], $\xi_{FM} = 3.45/(T/T_C - 1)^{1/2}$, is much less than the SP size, $2R$, which causes strong SP localization due to the significant energy shift of the nearest equivalent SP positions within the lattice (so-called "static destruction of the band"[49,51]). At lower temperature, once $\xi_{FM}$ exceeds the SP size, energy levels equilibrate within the length scale of $\xi_{FM}$, which causes SP delocalization. A simple estimate shows that $\xi_{FM}$ becomes equal to $2R$ at 77 K, which agrees well with the experiment. Thus, a crossover at about 80 K (Figure 2, Figure 3 and Figure 6) marks a fundamental change from localized to itinerant SP behaviour in $UGe_2$. A similar crossover in SP dynamics is found in correlated metallic $Cd_2Re_2O_7$, in which the classical Boltzmann distribution of the spectral weights breaks down to leave a uniform distribution within a narrow SP band[59]. In $UGe_2$, the small difference in amplitudes below 80 K (Figure 3, also found in Ref. 84) is due to the difference in Zeeman energy between the spin-up and spin-down states in the ferromagnetic environment. What is more fundamental is that the amplitudes or spectral weights of both lines are *temperature independent* below about 80 K (Figure 3), which clearly indicates temperature independent populations of the spin-up and spin-down states — which in turn is inconsistent with a localized SP but rather signifies its itinerant nature. Thus, our model suggests that the SP, captured by the muon, stays localized and is protected from spin exchange above 80 K; however, below 80 K it remains localized but undergoes active spin exchange with free itinerant SP. The increased length scale of the FM fluctuations ensures the remarkable itinerancy of the SP.

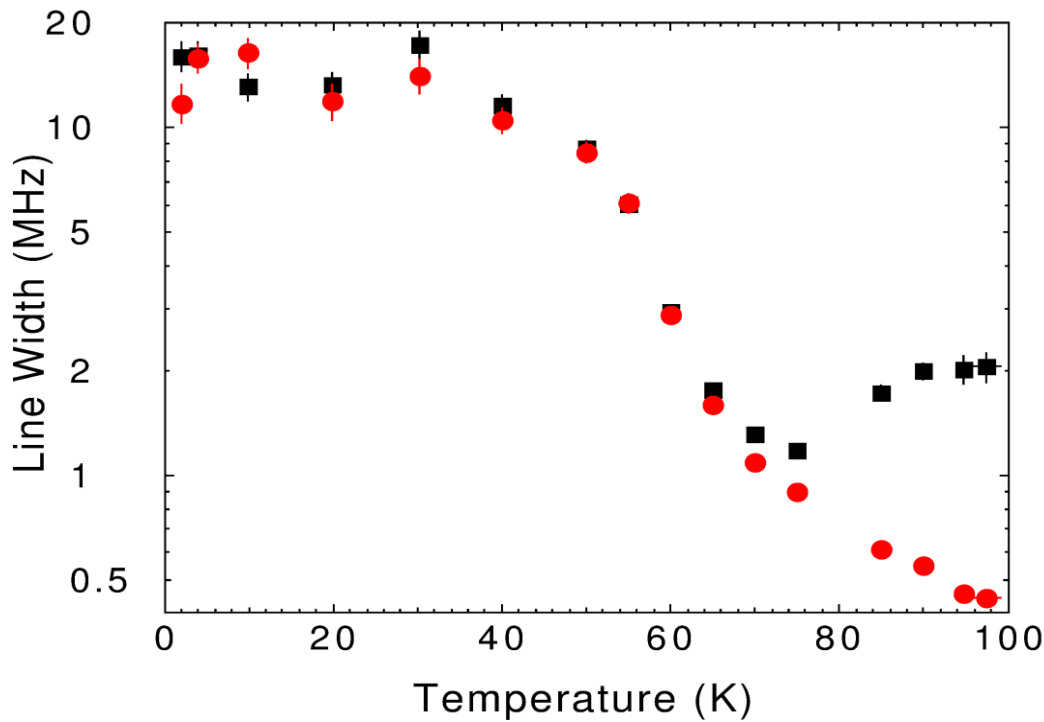

**Figure 6** Temperature dependences of the SP line widths of the signals with higher (red circles) and lower (black squares) frequencies in UGe$_2$ at 100 K in a magnetic field of $H = 1$ T. Note the qualitative change around 80 K.

Another remarkable feature which accompanies the onset of the SP itinerancy at 80 K is that the majority of the spin states at high temperature (red circles in Figure 3) becomes a minority of spin states at low temperature (with the corresponding changeover of the black squares, Figure 3). The distribution of the spin states is controlled by the Hund's rule, which ensures that the SP electron's spin ($s$) in UGe$_2$ will be parallel to electron spins of the unfilled 5$f$ orbital, similar to that in unfilled 3$d$ orbital in MnSi[66] or CdCr$_2$Se$_4$[64] and in marked contrast to completely filled 4$f$ orbital in Eu-based materials[60,63] (see Ref. 64 for a discussion on the Hund's rule control of the spin states within the SP). In UGe$_2$, the orbital magnetic moment of the U ion is twice as large as, and antiparallel to, that of the U ion spin[88], so that the SP electron spin ($s$) in the majority spin state finds itself to be opposite to the aggregate spin ($S$) of the SP at high temperature. The remarkable crossover at 80 K indicates that the majority of spin states prefer the orientation of the SP electron spin ($s$) to be parallel to $S$ at low temperature. As the SP electron spin (*s*) couples to spin moments of U ions, and not to their orbital moments, this parallel orientation reveals the departure from Hund's rule, with a corresponding increase of the Hund energy. Such an increase is very unlikely unless it is compensated by a decrease of some other energy by at least as much. The compensation comes from the significant decrease in the kinetic energy of the SP in the process of its delocalization, which presents another indication of the SP itinerancy below 80 K. A crossover from the antiparallel orientation of *s* (with respect to *S*) to a parallel orientation within the SP, occurs for the fundamental reason of greatly reducing spin scattering, allowing for much higher SP mobility within the local FM environment and hence, within the FM fluctuations — which occurs once the size of the fluctuations exceed the SP size at lower temperature.

Thus, at high temperature the majority of the localized SP states have their electron spin (*s*) antiparallel to their aggregate spin (*S*) which is dominated by the orbital moments of U ions. Quite in contrast, at low temperature the majority of SP states are itinerant SP with their spin *S* still dominated by the orbital moments of U ions but their electron spin (*s*) parallel to *S*. A significant consequence of the crossover at 80 K is that at low temperature, magnetic moments of itinerant SP are parallel to those of U ions, once the latter order along the *a*-axis below $T_c$, which suppresses spin-flip scattering events, providing greater SP mobility.

However, the remarkable itinerancy acquired by SP below 80 K does not mean they form a band at 80 K. Instead, spin polaron transport in UGe$_2$ just below 80 K is determined by hopping (diffusive) dynamics similar to the hopping dynamics of muon or Mu states in the regime dominated by "static destruction of the band" when dynamic fluctuations of the environment (phonons in the case of Mu) makes such hopping dynamics possible[49,51,91,92]. The latter regime is characterized by spatially inhomogeneous tunnelling dynamics when the small size of the particle bandwidth compared to all other energy parameters of the crystal (specifically, the typical difference between energy levels at adjacent tunnelling sites) restricts particle tunnelling dynamics to undisturbed regions and causes strong localization outside such regions. Likewise below 80 K, spin polarons in UGe$_2$ acquire itinerancy in a *restricted space* within the size of the FM fluctuation ($\xi_{FM}$) once $\xi_{FM} > 2R$ (see above). At lower temperature, where $\xi_{FM}$ exceeds the average distance between SP, polarons are no longer isolated but rather experience strong inelastic mutual spin-flip scattering, which again disables their coherent dynamics.

Evidence for the appearance at low temperature of spin-polarized itinerant carriers with magnetic moments of about $0.02\mu_B$ per U atom is present in several experiments on UGe$_2$, including measurements of magnetic entropy[34], muon spin relaxation in zero magnetic field[42], neutron measurements and magnetic susceptibility[93]. Optical studies[87] indicate a suppression of spin-flip scattering below $T_c$ with a significant increase of itinerant carrier effective mass to about $25m_0$. All of these facts indicate the formation of a spin-polarized band composed of heavy carriers at low temperature in UGe$_2$. We argue that this band is a SP band. Such a SP band will form near $T_c$ when FM fluctuations extend over the entire crystal and SP become completely delocalized. However, mutual spin-flip scattering between SP, depending on the SP size and aggregate spin, makes their dynamics incoherent.

In fact, coherent SP transport is only possible in a spin-polarized band where spin-flip scattering is suppressed due to the absence of SP with the opposite spin state. In a FM state, such coherent transport of SP becomes possible due to splitting of the majority and minority spin subbands. This splitting may be viewed as Zeeman splitting due to spontaneous magnetization. As a result, a characteristic sharp double peak structure is formed in the density of states (DOS) at the Fermi level. In UGe$_2$, such splitting ($\Delta$) is not large enough to leave the minority spin band empty[4,94]. The crucial point here is that $\Delta << W_{SP}$, where $W_{SP}$ is the SP bandwidth. Therefore, the Fermi surface contains both majority and minority spin sheets. Position of the Fermi level determines the spin polarization, $P = (n\downarrow - n\uparrow)/(n\downarrow + n\uparrow)$, which is essentially nonzero at low temperature[95]. This remarkable feature not only ensures coherent band dynamics of SP, but is also of a fundamental importance to the superconducting state: NQR measurements have revealed that the spin-up band is gapped but the spin-down band remains gapless, thereby indicating the unconventional nature of SC in UGe$_2$[95].

The primary condition for the coherent band dynamics of SP is suppression of spin-flip scattering on the Fermi level when $\Delta(T) > 2k_BT$, where $2k_BT$ is the thermal broadening of the Fermi function. We find the characteristic temperature of coherent SP band formation from

$$2k_B T_{SPB} \cong \Delta(T) \qquad (3)$$

In FM metals, spin splitting follows the bulk magnetization[96]. According to neutron studies of the temperature dependence of the magnetic moment in UGe$_2$[35],

$$\Delta(T) \cong \Delta_0 \, (1 - T/T_C)^{0.3}, \qquad (4)$$

where $\Delta_0$ is the spin splitting at $T = 0$. On the other hand, following[96]

$$\Delta_0 = P \, W_{SP} \qquad (5)$$

according to our data (Figure 3), the difference between majority and minority spin SP states, $P = 0.12$, which is close to the spin polarization of charge carriers found by NQR[95] at elevated pressures (1.17 and 1.2 Gpa): $P = 0.14$.

We determine the SP bandwidth from[47]

$$W_{SP} = z \, \hbar^2/(m_{SP} a^2), \qquad (6)$$

where $z$ is a coordination number, $m_{MP}$ is the SP mass and $a$ is the lattice constant. For a crude estimate we use UGe$_2$ values[34] $z = 4$, $m_{SP} = 25$ and $a = 0.4$ nm to get $W_{SP} \approx 70$ meV, $\Delta_0 \approx 8$ meV and to confirm that $\Delta << W_{SP}$. Then using (3) and (4) we arrive at the relation $T_{SPB} = 46 \, (1 - T_{SPB}/T_C)^{0.3}$ which, solved numerically, yields $T_{SPB} = 34$ K, in close agreement with a characteristic crossover temperature $T^* = 30$ K[35,34].

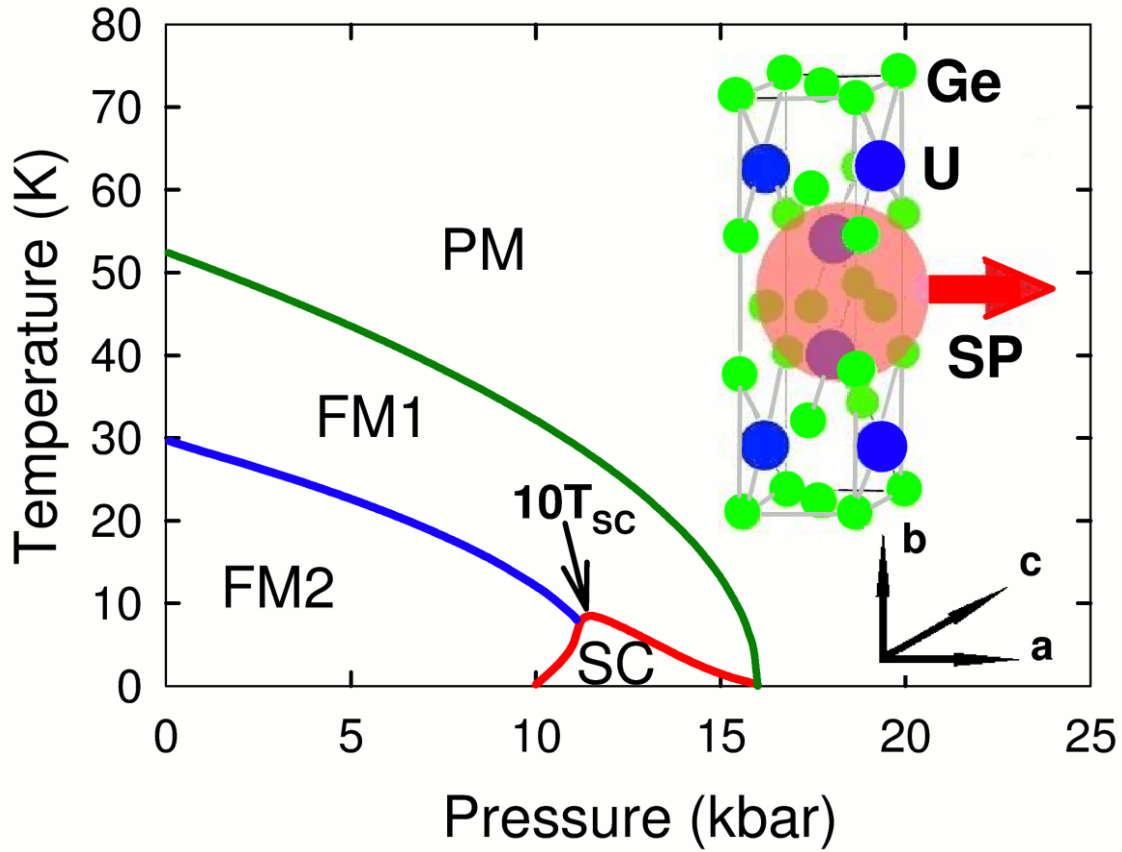

**Figure 7** The temperature-pressure phase diagram of UGe$_2$. The blue line ($T^*$) separates the weakly polarized magnetic phase (FM1) from the strongly polarized magnetic phase (FM2). The superconducting phase (SC) is confined within the red line (note that the $T_{SC}$ values are scaled by a factor of 10). The green line separates the paramagnetic (PM) phase from the ferromagnetic phase (FM1). The upper right inset shows the orthorhombic unit cell of UGe$_2$, composed of the U atoms (blue circles), Ge atoms (green circles) and a spin polaron (SP, pink circle) that confines two U atoms. The lower right inset shows the *a,b,c* axes of the UGe$_2$ unit cell. The SP propagates along easy axis *a*, which is also its spin direction (see text).

In UGe$_2$, this weak first-order transition (the crossover at ambient pressure) at $T^*$ separates the high-temperature (high pressure) weakly polarized phase (FM1) from the low-temperature (low pressure) strongly polarized phase (FM2)[35] (see Figure 7) and shows up as anomalies in resistivity[4,37], heat capacity[37,34], magnetization[97] and magnetoresistivity[34]. What is more fundamental is that this transition shows up as a distinct change in the Fermi surface as revealed by dHvA measurements[38-40]. The nature of the transition at $T^*$ is not clear at the moment, but experimental results indicate it to be closely related to superconductivity in UGe$_2$[3,4,35]. It has been suggested that this transition is due to charge or spin density wave(s) or both. However, neutron studies do not detect any such phases[35]. An alternative suggestion, which is also supported by our data, is that there is a first-order Stoner-type phase transition in the spin magnetization due to a sharp double peak in the density of states near $E_F$[94]. In the latter scenario, both the FM1-FM2 and SC transitions are driven in the FM phase by tuning the majority spin

Fermi level through one of two peaks in the density of states. The tuning parameter is the magnetization, which changes the topology of the Fermi surface for different spin species[94]. Magnetization measurements[97] do indicate that the FM1-FM2 transition occurs at a particular spin splitting between the majority and minority spin bands, since the Fermi level passes through a sharp peak in the DOS for one spin species. A qualitative change in DOS at the FM1-FM2 transition is also supported by dHvA[38-40] and Hall[41] measurements, although the nature of the spin bands and spin species is not discussed. We suggest that such a band is made of SP and the double-peak DOS is due to spin splitting of SP band, as discussed above. We argue that the FM1-FM2 transition may be a Stoner-type transition where coherent SP band dynamics sets in at $T^* = T_{SPB}$. This is consistent with an increase of the easy axis magnetic moment per U ion at $T^*$ in the low temperature FM2 phase, with respect to that in FM1[97].

Further support for coherent SP band formation at $T^*$ comes from the measurements of the local magnetic field shift on the muon ($\Delta B$) with respect to external magnetic field $B_0$ (Figure 8). The local magnetic field on the muon ($B$) includes all contributions from the magnetic environment (from both local moments and itinerant species) but excludes a contribution from the hyperfine field of the electron that belongs to the SP captured by the muon. In other words, $\Delta B$ presents the magnetic field shift on the muon as if this muon stays bare and does not capture a SP, similar to the shift measured in EuS[60]. It is determined as $\Delta B = B_0 - 2\pi (v_1+v_2)/2\gamma_\mu$, where $v_1$ and $v_2$ are the two frequencies presented in Figure 4, as they appear symmetrically split by $\pm A/2$ about the hypothetical bare muon frequency: $v_{1,2} = \pm A/2 + 2\pi (B+B_0)/\gamma_\mu$[60,75,78]. At high temperature, $\Delta B$ follows the bulk magnetization (measured in the same sample using a SQUID magnetometer). In this case, the itinerant species' contribution to the local field is small enough that both techniques result in similar measurements, mainly contributions from the local magnetic moments of U ions. However, below $T_c$ where FM fluctuations extend over the entire crystal, spin polarons become completely delocalized and tend to screen the magnetic moment of a single localized SP captured by the muon, thereby reducing the magnetic field on the muon, which causes the deviation of $\Delta B$ from the bulk magnetization (Figure 8). This effect is analogous to Kondo screening in the single magnetic impurity Kondo problem[98]. The difference is that the canonical Kondo effect involves free electrons, while in our case the screening cloud consists entirely of itinerant spin polarons. In fact, deviation of $\Delta B$ from the bulk magnetization starts below $T_c$, which sets up a nonzero spontaneous internal magnetic field that determines the nonzero Zeeman splitting ($\Delta$) between SP spin-up and spin-down species. However, this contribution stays small (although significant) above $T^*$ once $\Delta(T) < k_B T^*$. A qualitative change sets in below $T^* = 30$ K (Figure 8) once $\Delta(T)$ exceeds this temperature (see above discussion). Formation of a coherent SP band causes effective Kondo-like screening of the muon-captured SP by itinerant spin polarons, which determines a strong reduction of $\Delta B$ below $T^*$, while the bulk magnetization (performed *via* SQUID measurements) continue to increase below $T^*$.

We note that this Kondo-like screening of the muon-captured SP is only possible within the SP system.

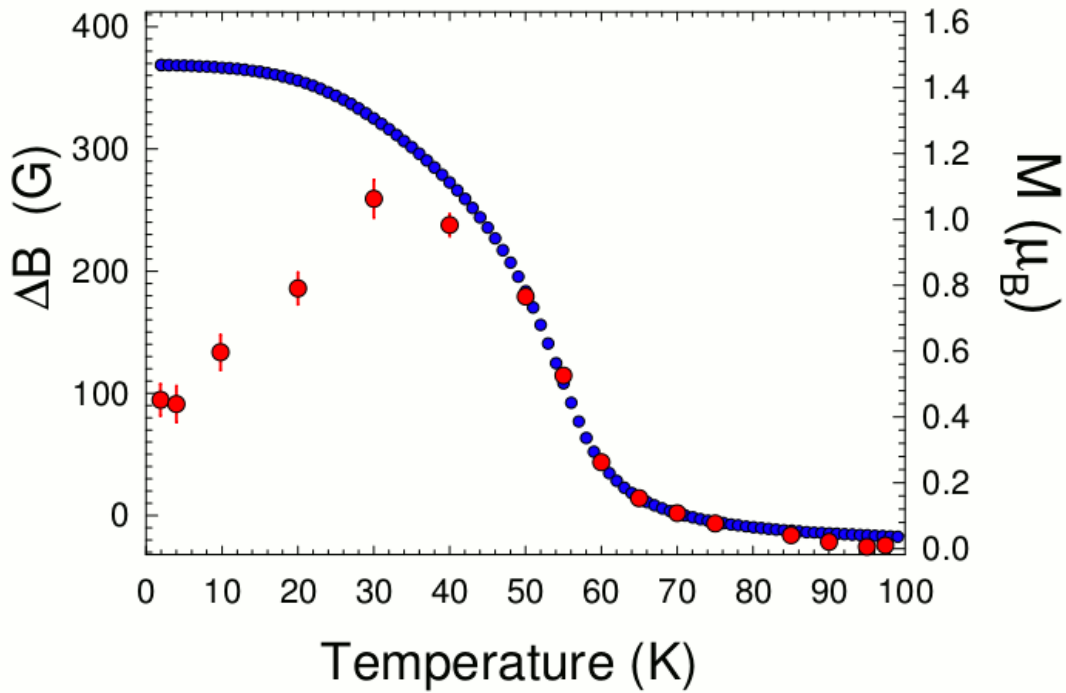

**Figure 8** Temperature dependence of the magnetic field shift detected by the muon (red circles) and bulk magnetization measurements (blue circles, presented as magnetic moment per U ion) in UGe$_2$. All data were collected with a magnetic field of $H = 1$ T externally applied parallel to the easy axis, *a*.

Thus, a non-monotonic behaviour of the magnetic field shift on the muon with its maximum at $T^*$ (displayed in Figure 8) might be explained by formation of a Kondo screening cloud made of SP within the spin-polaron band. The characteristic screening length scale of such a Kondo cloud can be estimated[99] as $\xi = \hbar v_F/k_B T_K$, where $v_F$ is the Fermi velocity of a SP within the SP band. Setting $T_K = T_{SPB} = T^*$ and getting $v_F \approx 3\times 10^6$ cm/s (which is about 100 times less than that for electrons in a metal being renormalized by the effective mass) from the SP band width, $W_{SP} \approx 70$ meV, with the SP mass, $m_{SP} = 25 m_0$, we arrive at $\xi \approx 0.7\times 10^{-6}$ cm. For comparison, depending on the Kondo temperature, the Kondo cloud in a canonical system of a single magnetic impurity in a simple metal may have a significant extension of $\sim 10^{-4}$ cm[100]. Although the characteristic size of a Kondo cloud made of spin polarons is about 100 times smaller than that in a canonical system, it is much larger than the SP size, which ensures effective screening within the spin polaron band.

Formation of such a narrow SP band should be directly relevant to a large electron mass enhancement and critical spin fluctuations as are inferred from optical conductivity measurements in UGe$_2$[87] and several other strongly correlated materials including UPd$_2$Al$_3$[101], UPt$_3$[101], Cd$_2$Re$_2$O$_7$[102], MnSi[103] and UBe$_{13}$[104]. At low temperature, all of these materials exhibit a narrow peak separated by a hybridization gap from a broad mid-infrared excitation feature in the frequency-dependent optical conductivity, which are associated with a low-spectral-weight, low-energy "heavy fermion" Drude peak and a high-spectral-weight, high-energy interband transition, respectively. In the latter three materials, observation of itinerant spin polarons[59,66,85] appears to be associated with the low-energy and mid-infrared features detected by optical conductivity[102-104]. Furthermore, in all of these materials enhancement of the effective mass (to several dozens of $m_0$ at low $T$) and suppression of the scattering rate mainly occur below a characteristic energy, consistent with the formation of a SP band.

Formation of a spin polaron band offers a straightforward explanation of the remarkable magnetoresistance (MR) in $UGe_2$[34]. Both MR and magnetostriction[105] indicate that the carrier number is a strong function of magnetic field. The remarkable sensitivity of the electron transport to the magnetization (an order of magnitude stronger than in hole-doped manganites) shows that carrier localization into SP and its release by a magnetic field may be a missing key ingredient of HF models. A SP model[106,107] predicts the MR to be dependent on the carrier density through formation of SP and carrier release from SP. The recent observation of SP in magnetoresistive $Lu_2V_2O_7$[83] supports this model. High magnetic field destroys the SP because in high $B$ the spins are already polarized, so that the exchange coupling of the carrier with these spins offers no energy advantage to compensate the increase in kinetic energy that occurs due to localization. Application of a magnetic field releases the carrier from SP into the conduction band — a process that not only explains the huge negative MR and its strong anisotropy, but also reveals the reason why carrier number is a strong function of magnetic field in $UGe_2$[41]. This effect may be relevant to earlier studies on suppression of heavy fermions by high magnetic field[108-110].

Formation of the spin polaron band may provide an explanation for a list of mysteries of $UGe_2$ as well as other HF and strongly correlated materials, such as itinerant versus localized carriers in a duality problem when local moments acquire itinerancy, "small" Fermi surface versus "large" Fermi surface including both conduction electrons and local spins, huge anisotropic magnetoresistance, Fermi surface reconstruction at $T^*$ and the nature of the mysterious FM transition within the FM phase. Emergence of such SP bands might be a general phenomenon in HF systems[85,111]. Furthermore, formation of spin bipolarons[59,112] — a pairing within the SP band — may be a missing ingredient for magnetically mediated spin-triplet SC models, including those in HF materials, FM materials or both. Finally, a spin polaron may serve as a *composite quasiparticle* — a heavy fermion — in HF materials.

This work was supported by the Natural Sciences and Engineering Research Council of Canada and the U.S. Department of Energy, Basic Energy Sciences, Division of Materials Science and Engineering (Award # DE-SC0001769).